# SPACE CHARGE MEASUREMENTS WITH A HIGH INTENSITY BUNCH AT THE FERMILAB MAIN INJECTOR*

K. Seiya[#], B. Chase, J. Dey, P. Joireman, I. Kourbanis, Fermilab, Batavia, IL 60439, U.S.A.
A. Yagodnitsyna, Novosibirsk State University, Russia (Fermilab summer student)

*Abstract*

For Project X, the Fermilab Main Injector will be required to operate with 3 times higher bunch intensity. The plan to study the space charge effects at the injection energy with intense bunches will be discussed.

## INTRODUCTION

A multi-MW proton facility has been established as a critical need for the U.S. HEP program by HEPAP and P5 [1]. Utilization of the Main Injector (MI) as a high intensity proton source capable of delivering in excess of 2 MW beam power will require a factor of three increase in bunch intensity compared to current operations. Instabilities associated with beam loading, space charge, and electron cloud effects are common issues for high intensity proton machines. The MI intensities for current operations and Project X are listed in Table 1.

The MI provides proton beams for Fermilab's Tevatron Proton-Antiproton Collider and MINOS neutrino experiments. The proposed 2MW proton facility, Project X, utilizes both the Recycler (RR) and the MI. The RR will be reconfigured as a proton accumulator and injector to realize the factor 3 bunch intensity increase in the MI. Since the energy in the RR and the MI at injection will be 6-8 GeV, which is relatively low, space charge effects will be significant and need to be studied.

Studies based on the formation of high intensity bunches in the MI will guide the design and fabrication of the RF cavities and space-charge mitigation devices required for 2 MW operation of the MI. It is possible to create the higher bunch intensities required in the MI using a coalescing technique that has been successfully developed at Fermilab [2].

This paper will discuss a 5 bunch coalescing scheme at 8 GeV which will produce $2.5\times10^{11}$ protons in one bunch. Bunch stretching will be added to the coalescing process [3]. The required RF parameters were optimized with longitudinal simulations. The beam studies, that have a goal of 85% coalescing efficiency, were started in June 2010 [4].

## COALESCING OPERATION IN THE MI

Coalescing is a process in which several bunches are concentrated into one bunch by using lower frequency RF. Four 8 GeV antiproton batches, with each batch consisting of 5 bunches, are injected from the RR. These batches are accelerated to 150 GeV and each batch is concentrated into one bunch. The wall current monitor

Table 1: Beam power in the MI

|  | **Current operation** | **Project X** |
|---|---|---|
| Beam power | 400 kW | 2 MW |
| Total intensity | $4.0\times10^{13}$ | $1.6\times10^{14}$ |
| Number of bunches | 492 | 548 |
| Intensity per bunch | $1.0\times10^{11}$ | $3.0\times10^{11}$ |
| MI cycle rate | 2.2 sec | 1.4 sec |

signals during coalescing operation are shown in Fig.1. In standard operations, the following four steps are taken, (also shown in Fig. 2):

Step I: Five to seven 150 GeV bunches are circulating, captured in a 53 MHz, 0.5 MV RF system. The bunch shape matches the 0.5 MV RF bucket.

Step II: The voltage is decreased to 40 kV, and the bunches start rotating in phase space. After a quarter of a synchrotron period, the momentum spread is at a minimum.

Step III: The 53 MHz RF voltage is turned off. At the same time, a 2.5 MHz RF system with a voltage of 52 kV is brought on. The bunches start rotating in the 2.5 MHz RF bucket. 15 kV of 5.0 MHz RF is used as a second harmonic. After a quarter of a synchrotron period, the bunch length becomes a minimum.

Step IV: The 2.5 MHz RF voltage is turned off. At the same time, the 53 MHz RF voltage is brought to 0.5 MV and captures the coalesced bunches (recapture).

The efficiency of coalescing is 88% and 94% for each proton and antiproton bunch in current operations.

___
*Work supported by Fermilab Research Alliance, LLC under Contract No. DE-AC02-07CH11359 with the United States Department of Energy
[#]kiyomi@fnal.gov

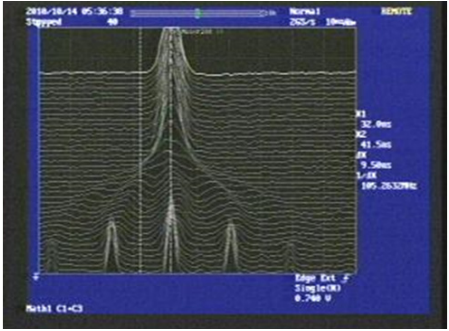

Figure 1: Mountain range plot with the wall current monitor signal during anti proton coalescing every 300 turns for 40 traces in the MI.

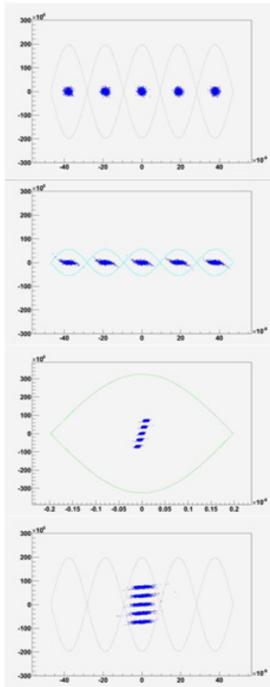

Figure 2: Longitudinal phase space on the coalescing steps from I to IV.

## SIMULATION RESULTS OF 8 GEV COALESCING

### 8 GeV coalescing with existing RF cavities

Using the same 2.5 MHz voltage as is used at 150 GeV, a coalescing simulation was carried out at 8 GeV. Figure 3 shows the longitudinal phase space after coalescing at 150 GeV and at 8 GeV. The bucket height is a factor of 9 lower than the 150 GeV case. In order to minimize the bunch length after the recapture the RF voltage needs to be higher, or the emittance before coalescing smaller. The existing 2.5 MHz MI RF cavities can produce a 75 kV maximum voltage, which is too small a voltage for 85% efficiency after recapture.

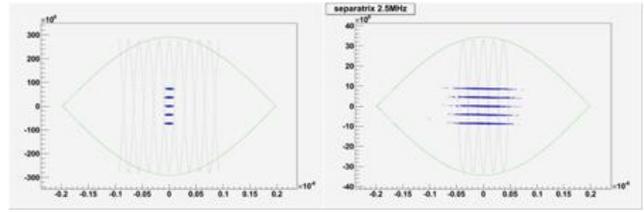

Figure 3: Longitudinal phase space after coalescing at 150GeV (left) and 8GeV (right).

### Estimation of the required emittance before coalescing

The required emittance was estimated from simulation results. The coalescing efficiency was simulated with a bunch length before coalescing of +/-1 nsec and 9 nsec and an energy spread from +/- 1 MeV to 10 MeV as shown in Fig. 4. The results show that the bunch length before coalescing does not affect the efficiency, but the energy spread should be less than +/- 3MeV for 85% efficiency.

In current operations the minimum energy spread at injection is +/-6 MeV, using bunch rotation at Booster extraction. In simulations and beam studies, reducing the 53 MHz RF voltage to 20 kV adiabatically at injection reduces the energy spread of the beam, with 0.1 eV s emittance, to +/- 4.6 MeV. As the RF bucket is filled with particles it is not possible to lower the voltage below 20 kV. Therefore, both these schemes are not able to achieve an energy spread of +/- 3MeV.

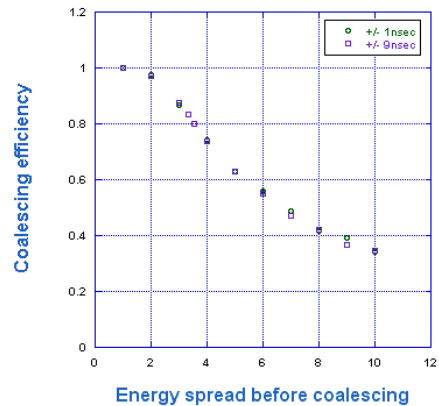

Figure 4: Coalescing efficiency with 75 kV on the 2.5MHz RF for +/-1nsec and 9nsec bunch length.

### Bunch stretching on the unstable fixed point

In order to reduce the energy spread, bunch stretching was applied to the simulation using the maximum voltage of 75 kV. When 5 bunches rotate by about 3/8 of a synchrotron period, the RF phase is jumped by 180 degrees. The 5 bunches start following the separatrix of the RF bucket, as shown in Fig. 5. The RF phase is jumped back to 0 degree before the bunches are affected by nonlinear effects from the RF. The bunches continue to rotate until the total bunch length becomes a minimum.

Figure 6 shows the longitudinal phase space at recapture. The length of the rotation time and phase jump period were optimised for energy spreads from +/- 1 MeV to +/-8 MeV with a bunch length of +/- 9 nsec. The maximized coalescing efficiency at each energy spread is shown in Fig. 7. Less than 5 MeV energy spread can produce 85% coalescing efficiency by reducing the RF voltage adiabatically.

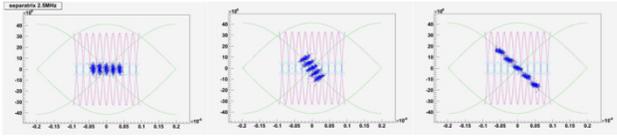

Figure 5: Longitudinal phase space during bunch stretching.

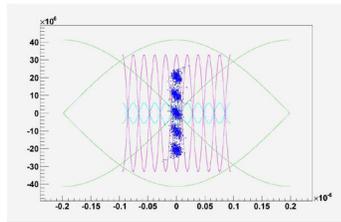

Figure 6: Longitudinal phase space at recapture with bunch stretching.

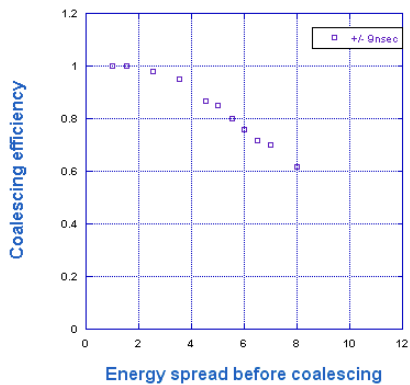

Figure 7: Coalescing efficiency with 75 kV on the 2.5MHz RF for +/- 9nsec bunch length with bunch stretching.

*Required RF parameters*

The optimized RF parameters for a beam with 0.1 eV s are given below and shown in Fig. 8.

Step I: Five bunches, at an energy of 8 GeV, are circulating, captured by a 53 MHz RF system with a voltage of 1.0 MV. The bunch shape matches the 1.0 MV RF bucket.

Step II: The voltage is decreased to 30 kV adiabatically in 0.2sec.

Step III: The 53 MHz RF voltage is changed to 0 V. At the same time, a 2.5 MHz RF system with a voltage of 75kV is brought on. The bunches start rotating in the 2.5 MHz RF bucket. 23 kV of 5.0 MHz RF voltage is used as a second harmonic.

Step IV: After 7.7 msec the RF phase of 2.5 MHz is jumped by 180 degrees. The 5 bunches start following the separatrix of the RF bucket for 2.5 msec and then the RF phase is jumped back to 0 degrees.

Step V: The bunches continue to rotate for another 17.4 msec until the total bunch length becomes minimum.

All RF parameters are producible with the existing high level and low level RF (LLRF) systems except the time resolution for the phase jump and voltage change. The LLRF code will be modified to improve the timing resolution from 1.33 msec to the required 60 μsec.

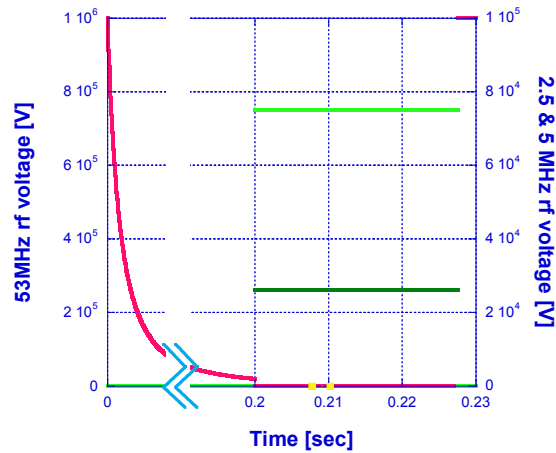

Figure 8: Optimized RF parameters for the bunch with 0.1 eV s of longitudinal emittance. 53 MHz RF voltage (magenta), 2.5 MHz (light green) and 5.0 MHz (green) RF voltage and timing for the phase jump (yellow).

## SUMMARY

Plans for generating high intensity bunches in the MI for the future project, Project X, have been studied. The beam studies were started in June 2010. Using coalescing and bunch stretching can produce $2.5 \times 10^{11}$ protons in one bunch at 8 GeV. The required RF parameters were optimized with longitudinal beam simulations. LLRF code is going to be modified to improve the timing resolution for the required time resolution.